\documentclass[11pt, onecolumn]{IEEEtran}

\usepackage{graphicx} % for pdf, bitmapped graphics files
\usepackage{amsmath} % assumes amsmath package installed
\usepackage{amssymb}  % assumes amsmath package installed
\usepackage{algorithm}
\usepackage{algorithmic}
\usepackage{cite}
\usepackage{subfig}

\newtheorem{theorem}{Theorem}
\newtheorem{lemma}{Lemma}
\newtheorem{defn}{Definition}
\newtheorem{corollary}{Corollary}
\newtheorem{remark}{Remark}

\IEEEoverridecommandlockouts

\newcommand{\prob}[1]{\mathbb{P}\left[#1\right]}
\newcommand{\expect}[1]{\mathbb{E}\left[#1\right]}

\newcommand{\pth}[1]{\left( #1 \right)}

\newcommand{\sth}[1]{\left\{ #1 \right\}}

\newcommand{\calC}{{\mathcal{C}}}
\newcommand{\calH}{{\mathcal{H}}}
\newcommand{\calI}{{\mathcal{I}}}
\newcommand{\calK}{{\mathcal{K}}}
\newcommand{\calM}{{\mathcal{M}}}
\newcommand{\calN}{{\mathcal{N}}}
\newcommand{\calP}{{\mathcal{P}}}
\newcommand{\calT}{{\mathcal{T}}}

\newcommand{\bbF}{{\mathbb{F}}}

\newcommand{\bfB}{{\mathbf{B}}}
\newcommand{\bfC}{{\mathbf{C}}}
\newcommand{\bfM}{{\mathbf{M}}}
\newcommand{\bfW}{{\mathbf{W}}}

\newcommand{\bfb}{{\mathbf{b}}}

\newcommand{\naturals}{\mathbb{N}}

\begin{document}

\title{Dynamic Distributed Storage for Scaling Blockchains} 

\author{Ravi Kiran Raman and Lav R.\ Varshney \thanks{The authors are with the University of Illinois at Urbana-Champaign.}}

\maketitle

%%%%%%%%%%%%%%%%%%%%%%%%%%%%%%%%%%%%%%%%%%%%%%%%%%%%%%%%%%%%%%%%%%%%%%%%%%%%%%%%
\begin{abstract}
Blockchain uses the idea of storing transaction data in the form of a distributed ledger wherein each node in the network stores a current copy of the sequence of transactions in the form of a hash chain. This requirement of storing the entire ledger incurs a high storage cost that grows undesirably large for high transaction rates and large networks. In this work we use the ideas of secret key sharing, private key encryption, and distributed storage to design a coding scheme such that each node stores only a part of the entire transaction thereby reducing the storage cost to a fraction of its original cost. When further using dynamic zone allocation, we show the coding scheme can also improve the integrity of the transaction data in the network over current schemes. Further, block validation (bitcoin mining) consumes a significant amount of energy as it is necessary to determine a hash value satisfying a specific set of constraints; we show that using dynamic distributed storage reduces these energy costs.
\end{abstract}
%%%%%%%%%%%%%%%%%%%%%%%%%%%%%%%%%%%%%%%%%%%%%%%%%%%%%%%%%%%%%%%%%%%%%%%%%%%%%%%%

\section{Introduction}

Blockchains are distributed, shared ledgers of transactions that reduce the friction in financial networks due to different intermediaries using different technology infrastructures, and even reduce the need for intermediaries to validate financial transactions. This has in turn led to the emergence of a new environment of business transactions and self-regulated cryptocurrencies such as bitcoin  \cite{BonneauMNKF2015, NarayananBFMG2016}. Owing to such favorable properties, blockchains are being adopted extensively outside cryptocurrencies in a variety of novel application domains \cite{TapscottT2016, Underwood2016}. In particular, blockchains have inspired novel innovations in medicine \cite{AzariaEVL2016}, supply chain management and global trade \cite{CaseyW2017}, and government services \cite{Swan2015}. Blockchains are expected to revolutionize the way financial/business transactions are done \cite{IansitiL2017}, for instance in the form of smart contracts \cite{KosbaMSWP2016}. In fact, streamlining infrastructure and removing redundant intermediaries creates the opportunity for significant efficiency gains.

The blockchain architecture however comes with an inherent weakness. The blockchain works on the premise that the entire ledger of transactions is stored in the form of a hash chain at every node, even though the transactions themselves are meaningless to the peers that are not party to the underlying transaction. Consequently, the individual nodes incur a significant, ever-increasing storage cost \cite{blockchaininfo}. Note that \emph{secure} storage may be much more costly than just raw hard drives, e.g.\ due to infrastructure and staffing costs.

In current practice, the most common technique to reduce the storage overload is to prune old transactions in the chain. However, this mechanism is not sustainable for blockchains that have to support a high arrival rate of transaction data. Hence the blockchain architecture is not scalable owing to the heavy storage requirements.

\begin{figure}[t]
	\centering
	\subfloat[Blockchain Size]{\label{fig:bitcoin_size}\includegraphics[scale = 0.34]{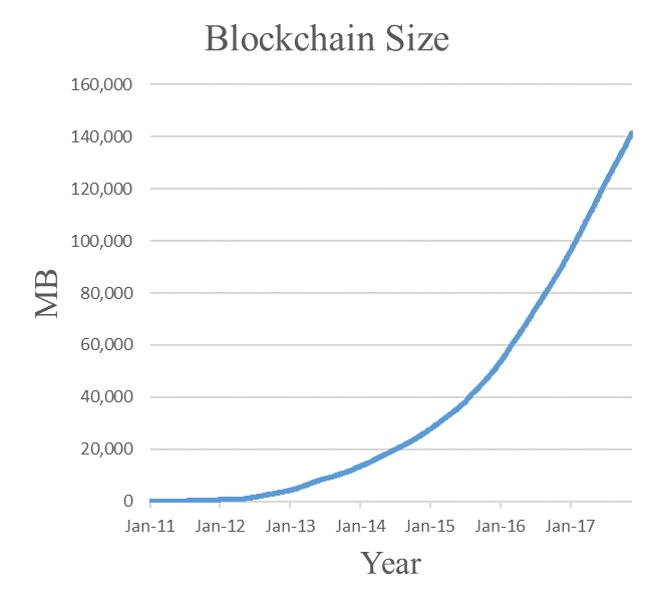}} 
	\subfloat[Average Number of Transactions per day]{\label{fig:bitcoin_transactions}\includegraphics[scale=0.34]{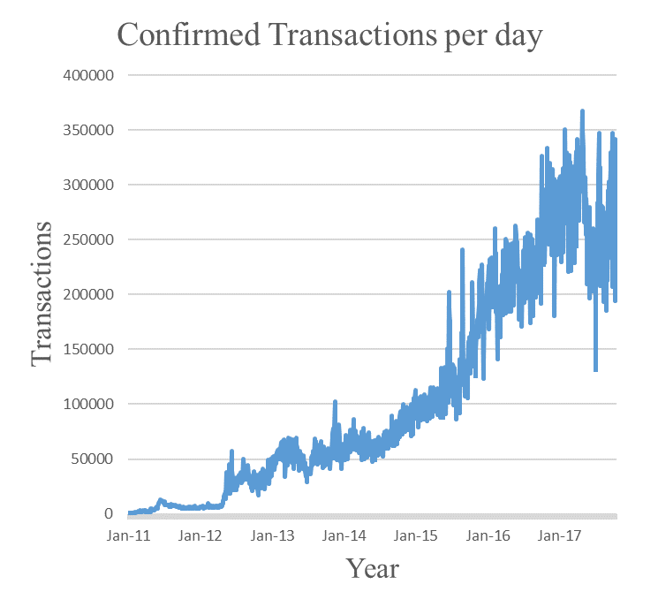}} 
	\subfloat[Hash-rate]{\label{fig:hash_rate}\includegraphics[scale=0.34]{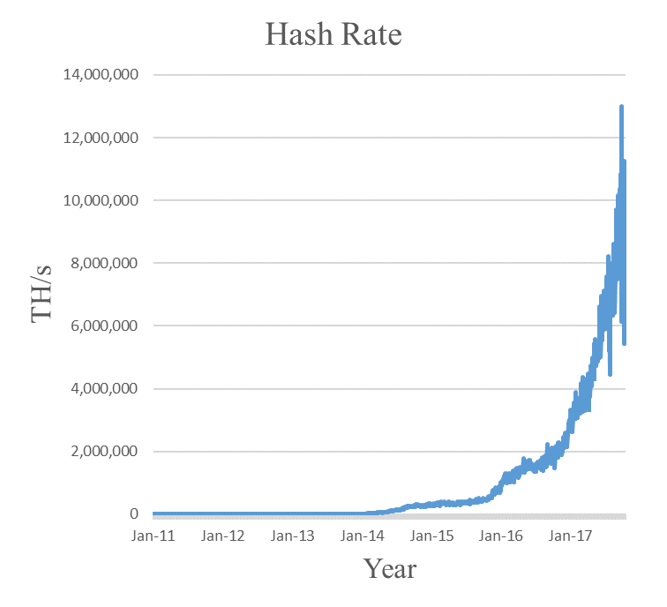}}
 	\caption{Increase in transactions, storage, and hash-rate in bitcoin.  Data obtained from \cite{blockchaininfo}.}
  	\label{fig:bitcoin_stats}
\end{figure}

For instance, consider the bitcoin network. Bitcoin currently serves an average of just under $3.5$ transactions per second \cite{CromanDEGJKMSSSSW2016}. This low number is owing to a variety of reasons including the economics involved in maintaining a high value for the bitcoin. However, even at this rate, this accounts for an average of $160$MB of storage per day \cite{blockchaininfo}, i.e., about $60$GB per year. The growth in storage and the number of transactions in the bitcoin network over time \cite{blockchaininfo} are shown in Figs.~\ref{fig:bitcoin_size} and \ref{fig:bitcoin_transactions}. The exponential growth in the size of the blockchain and the number of transactions to be served highlights the need for better storage techniques that can meet the surging demand.

SETL, an institutional payment and settlement infrastructure based on blockchain, claims to support 1 billion transactions per day. This is dwarfed by the Federal Reserve, which processes $14$ \emph{trillion} financial transactions per day. If cryptocurrencies are to become financial mainstays, they would need to be scaled by several orders. Specifically, they will eventually need to support to the order of about $2000$ transactions per second on average \cite{CromanDEGJKMSSSSW2016}. Note that this translates to an average storage cost of over $90$GB per day. This is to say nothing about uses for blockchain in global trade and commerce, healthcare, food and agriculture, and a wide variety of other industries. With storage cost expected to saturate soon due to the ending of Moore's Law, storage is emerging as a pressing concern with the large-scale adoption of blockchain.

This impending end to Moore's law not only implies saturated storage costs, but also a saturation of computational speeds. Block validation (mining) in bitcoin-like networks requires computing an appropriate hash value that caters to specified constraints and adding the data block to the blockchain. Notwithstanding new efforts such as \cite{VilimDK2016}, this process has been found to be computationally expensive and demands the use of high-end hardware and large volumes of energy. The hash rate, which is the number of hashes computed per second on the bitcoin network, is indicative of the total energy consumption by the network. The hash rate has been found to be increasing exponentially over time, as indicated in Fig.~\ref{fig:hash_rate}, as a result of both an increase in the rate of transactions, and an increase in the number of miners with powerful computing machines.

This staggering growth in hash computation results in a corresponding growth in energy consumption by the blockchain network. Recent independent studies have estimated that the global energy consumption of the bitcoin network is of the order of $700$MW \cite{Malmo2015, Fairley2017}. This amount of power is sufficient to power over $325$,$000$ homes and well over $5000$ times the power consumed per transaction by a credit card. Increasing demand for bitcoin in the absence of any updates to the mechanism of mining is thus projected to increase the consumption to the order of $14$GW by the year $2020$. The reducing energy efficiency improvement of mining hardware at a time when bitcoin value is rising rapidly, with a projected rapid increase the transaction rate further compounds the energy problem.

In addition to the unsustainable growth in energy consumption, this requirement of mining on availability of efficient computing resources and cheap electricity also establishes an imbalance in the decentralized mining requirement of bitcoin-like cryptocurrencies. In particular, miners tend to concentrate in geographic regions of the world where both the mentioned resources are available for cheap. This is again undesirable for cryptocurrencies such as bitcoin that rely fundamentally on the distributed nature of the mining operation to establish reliability. Thus, it is of interest to keep the mining process relatively cheap without compromising on data integrity. 

The current mechanism of bitcoin mining is referred to at large as a Proof of Work (PoW) method where the creation of a hash that adheres to the constraint set is identified as a proof of validation. This is typically done in a competitive environment with the first peer to mine earning a reward in terms of new bitcoins. The reason for using such a mechanism is two-fold. Firstly, it automates and controls the generation of new bitcoins in the system while also incentivizing miners to validate data blocks. Secondly, the creation of the hash constraints also enhances data integrity as a corruption to the block in turn requires the computationally expensive recomputation of the constrained hash values.

In this paper we do not consider the economics related to cryptocurencies through the mining process. Blockchain systems, cryptocurrencies in particular, however have explored the feasibility of reducing the energy demands by moving away from a competition-based mining process. Efforts to move to a Proof of Stake (PoS) model wherein peers are assigned the mining task in a deterministic fashion depending on the fraction of cryptocurrency (stake) they own. While the method is not yet foolproof, the removal of the competitive framework is expected to reduce the energy demands. In this work, we focus on energy reduction through an alternate scheme that guarantees a enhanced data integrity through information-theoretic immutability rather than a computational one in Sec.~\ref{sec:energy_redn}.

\subsection{Our Contributions}

In this paper we show that the storage costs in large-scale blockchain networks can be reduced through secure distributed storage codes. Specifically, when each peer stores a piece of the entire transaction instead of the entire block, the storage cost can be diminished to a fraction of the original. In Sec.~\ref{sec:coding_scheme}, we use a novel combination of Shamir's secret sharing scheme \cite{Shamir1979}, private key encryption, and distributed storage codes \cite{DimakisR2008}, inspired by \cite{Krawczyk1994}, to construct a coding scheme that distributes transaction data among subsets of peers. The scheme is shown to be optimal in storage up to small additive terms. We also show, using a combination of statistical and cryptographic security of the code, that the distributed storage loses an arbitrarily small amount of data integrity as compared to the conventional method.

Further, in Sec.~\ref{sec:zone_alloc}, using a dynamic zone allocation strategy among peers, we show that the integrity of the data can further be enhanced in the blockchain. We formulate the zone allocation problem as an interesting combinatorial problem of decomposing complete hypergraphs into $1$-factors. We design an allocation strategy that is order optimal in the time taken to ensure highest data integrity. We show that given enough transaction blocks to follow, such a system is secure from active adversaries.

Distributed storage schemes have been considered in the past in the form of information dispersal algorithms (IDA) \cite{Rabin1990, CachinT2005} and in the form of distributed storage codes \cite{DimakisR2008, RashmiSKR2009}. In particular \cite{CachinT2005} considers an information dispersal scheme that is secure from adaptive adversaries. We note that the coding scheme we define here is stronger than such methods as it handles active adversaries. Secure distributed storage codes with repair capabilities to protect against colluding eavesdroppers \cite{RawatKSV2014} and active adversaries \cite{PawarRR2011} have also been considered. The difference in the nature of attacks by adversaries calls for a new coding scheme in this paper.

In addition to the reduction in storage costs without compromising the integrity, in Sec.~\ref{sec:energy_redn}, we study the energy demands of our scheme in comparison to conventional bitcoin-like blockchains by comparing the computational complexity involved in establishing the blockchain. We show that our scheme creates a framework to reduce energy consumption through the use of simpler, easy to compute hash functions.

The enhanced integrity, reduced storage and energy costs do come at the expense of increased recovery and repair costs. In particular, as the data is stored as a secure distributed code, the recovery process is highly sensitive to denial of service attacks and the codes are not locally repairable, as elaborated in Sec.~\ref{sec:repair_recovery}.

Before getting into the construction and analysis of the codes, we first introduce a mathematical abstraction of the blockchain in Sec.~\ref{sec:model}, and give a brief introduction to the coding schemes that are used in this work in Sec.~\ref{sec:prelim}.

\section{System Model} \label{sec:model}

We now introduce a mathematical model of conventional blockchain systems. We first describe the method in which the distributed ledger is maintained in the peer network, emphasizing the roles of various nodes in the process of incorporating a new transaction in the ledger. Then, we introduce data corruption in blockchain systems and the nature of the active adversary considered in this work.

\subsection{Ledger Construction}

The blockchain comprises a connected peer-to-peer network of nodes, where nodes are placed into three primary categories based on functionality:
\begin{enumerate}
\item {\bf Clients:} nodes that invoke or are involved in a transaction, have the blocks validated by endorsers, and communicate them to the orderers.
\item {\bf Peers:} nodes that commit transactions and maintain a current version of the ledger. Peers may also adopt endorser roles.
\item {\bf Orderer:} nodes that communicate the transactions to the peers in chronological order to ensure consistency of the hash chain.
\end{enumerate}
Note that the classification highlighted here is only based on function, and individual nodes in the network can serve multiple roles.

\begin{figure}[t]
	\centering
	\includegraphics[scale = 0.5]{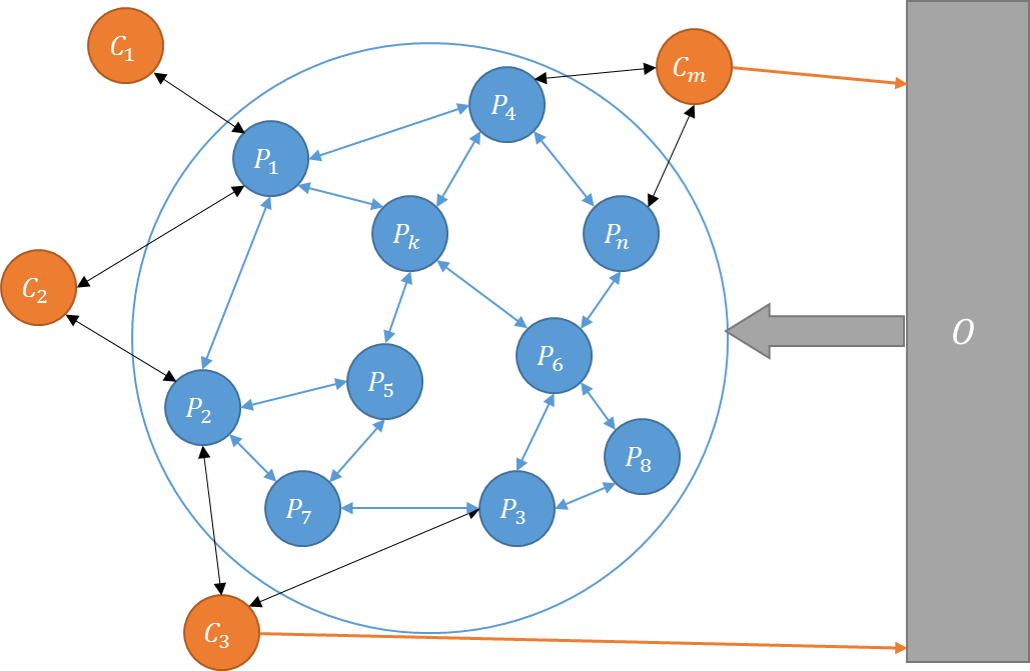}
 	\caption{Architecture of the Blockchain Network. Here the network is categorized by functional role into clients $C_i$, peers $P_i$, and orderers $O$. As mentioned earlier, the clients initialize transactions. Upon validation, the transactions are communicated to peers by orderers. The peers maintain an ordered copy of the ledger of transactions.}
  	\label{fig:blockchain_arch}
\end{figure}

The distributed ledger of the blockchain maintains a current copy of the sequence of transactions. A transaction is initiated by the participating clients and is verified by endorsers (select peers). Subsequently, the verified transaction is communicated to the orderer. The orderer then broadcasts the transaction blocks to the peers to store in the ledger. The nodes in the blockchain are as depicted in Fig. \ref{fig:blockchain_arch}. Here nodes $C_i$ are clients, $P_i$ are peers, and $O$ is the set of orderers in the system, categorized by function.

The ledger at each node is stored in the form of a (cryptographic) hash chain.
\begin{defn}
Let $\calM$ be a message space consisting of messages of arbitrary length. A \emph{cryptographic hash function} is a deterministic function $h: \calM \rightarrow \calH$, where $\calH$ is a set of fixed-length sequences called \emph{hash values}.
\end{defn}
Cryptographic hash functions additionally incorporate the following properties:
\begin{enumerate}
\item {\bf Computational ease:} To establish efficiency, the hash function should be easy to compute.
\item {\bf Pre-image resistance:} Given $H \in \calH$, it is computationally infeasible to find $M \in \calM$ such that $h(\calM) = H$.
\item {\bf Collision resistance:} It is computationally infeasible to find $M_1, M_2 \in \calM$ such that $h(M_2) = h(M_1)$. 
\item {\bf Sensitivity:} Minor changes in the input change the corresponding hash values significantly.
\end{enumerate}
Other application-specific properties are incorporated depending on need.

Let $\bfB_t$ be the $t$th transaction block and let $H_t$ be the hash value stored with the $(t+1)$th transaction, computed as $H_t = h(W_t)$, where $h(\cdot)$ is the hash function, and $W_t = (H_{t-1},\bfB_t)$ is the concatenation of $H_{t-1}$ and $\bfB_t$. Thus, the hash chain is stored as 
\[
(H_0,\bfB_1)-(H_1,\bfB_2)-\dots-(H_{t-1},\bfB_t).
\]

Let us assume that for all $t$, $\bfB_t \sim \text{Unif}(\bbF_q)$ and $H_t \in \bbF_p$, where $q,p\in\naturals$ and $\bbF_q, \bbF_p$ are finite fields of order $q$ and $p$ respectively. Thus, in the conventional implementation of the blockchain, the cost of storage per peer per transaction is 
\begin{equation} \label{eqn:original_cost}
\tilde{R}_s = \log_2 q + \log_2 p \text{ bits}.
\end{equation}

Transactions stored in the ledger may at a later point be recovered in order to validate claims or verify details of the past transaction by nodes that have read access to the data. Different implementations of the blockchain invoke different recovery mechanisms depending on the application. One such method is to use an authentication mechanism wherein select peers return the data stored in the ledger and the other peers validate (sign) the content. Depending on the application, one can envision varying the number of authorization checks necessary to validate the content.

For convenience, we restrict this work to one form of retrieval which broadly encompasses a wide class of recovery schemes. Specifically, we assume that in order to recover the $t$th transaction, each peer returns its copy of the transaction and the majority rule is applied to recover the block.

\subsection{Blockchain Security} \label{sec:sys_model_b}

There are two main forms of data security that we focus on.
\begin{enumerate}
\item {\bf Integrity:} transaction data stored by clients cannot be corrupted unless a majority of the peers are corrupted.
\item {\bf Confidentiality:} local information from individual peers does not reveal sensitive transaction information.
\end{enumerate}

Corruption of data in the blockchain requires corrupting a majority of the peers in the network to alter the data stored in the distributed, duplicated copy of the transaction ledger. Thus, the blockchain system automatically ensures a level of integrity in the transaction data.

Conventional blockchain systems such as bitcoin enforce additional constraints on the hash values to enhance data integrity. For instance, in the bitcoin network, each transaction block is appended with a nonce which is typically a string of zeros, such that the corresponding hash value satisfies a difficulty target i.e., is in a specified constraint set. The establishment of such difficulty targets in turn implies that computing a nonce to satisfy the hash constraints is computationally expensive. Thus data integrity can be tested by ensuring the hash values are consistent as it is computationally infeasible to alter the data.

The hash chain, even without the difficulty targets, offers a mechanism to ensure data integrity in the blockchain. Specifically, note that the sensitivity and pre-image resistance of the cryptographic hash function  ensures that any change to $H_{t-1}$ or $B_t$ would require recomputing $H_{t}$. Further, it is not computationally feasible to determine $W_{t}$ such that $h(W_t) = H_t$. 

Thus storing the ledger in the form of a hashchain ensures that corrupting a past transaction not only requires the client to corrupt at least half the set of peers to change the majority value, but also maintain a consistent hash chain following the corrupted transaction. That is, say a participating client wishes to alter transaction $B_1$ to $B_1'$. Let there be $T$ transactions in the ledger. Then, corrupting $B_1$ implies that the client would also have to replace $H_1$ with $H_1' = h(W_1')$ at the corrupted nodes. This creates a domino effect, in that all subsequent hashes must in turn be altered as well, to maintain integrity of the chain. This strengthens the integrity of the transaction data in blockchain systems.

Confidentiality of information is typically guaranteed in these systems through the use of private key encryption methods, where the key is shared with a select set of peers are authorized to view the contents of the transactions. Note however that in such implementations, a leak at a single node could lead to a complete disambiguation of the information.

\subsection{Active Adversary Model}

In this work, we explore the construction of a distributed storage coding scheme that ensures a heightened sense of confidentiality and integrity of the data, even when the hash functions are computationally inexpensive.

Let us assume that each transaction $B_t$ also has a corresponding \emph{access list}, which is the set of nodes that have permission to read and edit the content in $B_t$. Note that this is equivalent to holding a private key to decrypt the encrypted transaction data stored in the ledger.

In this work, we primarily focus on \emph{active adversaries} who alter a transaction content $B_t$ to a desired value $B_t'$. Let us explicitly define the semantic rules of a valid corruption for such an adversary. If a client corrupts a peer, then the client can
\begin{enumerate}
\item learn the contents stored in the peer;
\item alter block content only if it is in the access list of the corresponding block; and
\item alter hash values as long as chain integrity is preserved, i.e., an attacker cannot invalidate the transaction of another node in the process.
\end{enumerate}
The active adversary in our work is assumed to be aware of the contents of the hash chain and the block that it wishes to corrupt. We elaborate on the integrity of our coding scheme against such active adversaries. We also briefly elaborate on the data confidentiality guaranteed by our system against local information leaks.

Another typical attack of interest in such blockchain systems is the denial of service attack where an adversary corrupts a peer in the network to deny the requested service, which in this case is the data stored in the ledger. We also briefly describe the vulnerability of the system to denial of service attacks owing to the distributed storage. 

Before we describe the code construction, we first give a preliminary introduction to coding and encryption schemes that we use as the basis to build our coding scheme. 

\section{Preliminaries} \label{sec:prelim}

This work uses a private key encryption scheme with a novel combination of secret key sharing and distributed storage codes to store the transaction data and hash values. We now provide a brief introduction to these elements.

\subsection{Shamir's Secret Sharing}

Consider a secret $S \in \bbF_q$ that is to be shared with $n < q$ nodes such that any subset of size less than $k$ get no information regarding the secret upon collusion, while any subset of size at least $k$ get complete information. Shamir's $(k,n)$ secret sharing scheme \cite{Shamir1979} describes a method to explicitly construct such a code. All the arithmetic performed here is finite field arithmetic on $\bbF_q$.

Draw $a_i \stackrel{i.i.d}{\sim} \text{Unif}(\bbF_q)$, for $i \in [k-1]$ and set $a_0 = S$. Then, compute
\[
y_i = a_0 + a_1 x_i + a_2 x_i^2 + \dots + a_{k-1} x_i^{k-1}, \text{ for all } i \in [n],
\]
where $x_i = i$. Node $i \in [n]$ receives the share $y_i$. 

Since the values are computed according to a polynomial of order $k-1$, the coefficients of the polynomial can be uniquely determined only when we have access to at least $k$ points. Recovering the secret key involves polynomial interpolation of the $k$ shares to obtain the secret key (intercept). Thus, the secret can be recovered if and only if at least $k$ nodes collude.

In this work, we presume that each secret share is given by $(x_i,y_i)$, and that the unique abscissa values are chosen uniformly at random from $\bbF_q\backslash\{0\}$. That is, $\{x_i: i\in [n]\}$ are drawn uniformly at random without replacement from $\bbF_q\backslash\{0\}$. Then, given any $k-1$ shares and the secret, the final share uniformly likely in a set of size $q - k$.

It is worth noting that Shamir's scheme is minimal in storage as the size of each share is the same as the size of the secret key. Shamir's scheme however is not secure to active adversaries. In particular, by corrupting $n-k+1$ nodes, the secret can be completely altered.

Secret key sharing codes have been widely studied in the past \cite{McElieceS1981, KarninGH1983, HuangLKB2016}. In particular, it is known that Reed-Solomon codes can be adopted to define secret shares. Linear codes for minimal secret sharing have also been considered \cite{Massey1993}. In this work however, we restrict to Shamir's secret sharing scheme for simplicity.

\subsection{Data Encryption}

Shannon considered the question of perfect secrecy in cryptosystems from the standpoint of statistical security of encrypted data \cite{Shannon1949b}. There, he concluded that perfect secrecy required the use of keys drawn from a space as large as the message space. This is practically unusable as it is difficult to use and securely store such large key values. Thus practical cryptographic systems leverage computational limitations of an adversary to guarantee security over perfect statistical secrecy.

We define a notion of encryption that is slightly different from that used typically in cryptography. Consider a message $\bfM = (M_1,\dots,M_m) \in \calM$, drawn uniformly at random. Let $K \in \calK$ be a private key drawn uniformly at random.

\begin{defn}
Given message, key, and code spaces $\calM,\calK,\calC$ respectively, a \emph{private key encryption scheme} is a pair of functions $\Phi : \calM \times \calK \to \calC,~ \Psi : \calC \times \calK \to \calM$, such that for any $\bfM \in \calM$,
\[
\Phi(\bfM;K) = \bfC, \text{ such that, } \Psi(\bfC,K) = \bfM,
\]
and it is $\epsilon$-secure if it is statistically impossible to decrypt the codeword in the absence of the private key $K$ beyond a confidence of $\epsilon$ in the posterior probability. That is, if 
\begin{equation} \label{eqn:enc_given_code}
\max_{\bfM\in\calM, \bfC \in \calC} \prob{\Psi(\bfC,K) = M} \leq \epsilon.
\end{equation}
\end{defn}
The definition indicates that the encryption scheme is an invertible process and that it is statistically infeasible to decrypt the plaintext message beyond a degree of certainty. We know that given the codeword $\bfC$, decrypting the code is equivalent to identifying the chosen private key. In addition, from \eqref{eqn:enc_given_code}, we observe that the uncertainty in the message estimation is at least $\log_2 \pth{\tfrac{1}{\epsilon}}$. Thus,
\[
\log_2 \pth{\tfrac{1}{\epsilon}} \text{ bits} \leq H(\bfM \vert \bfC) \leq \log_2 |\calK| \text{ bits}.
\]

For convenience, we assume without loss of generality that the encrypted codewords are vectors of the same length as the message from an appropriate alphabet, i.e., $\bfC = (C_1,\dots,C_m)$.

Since we want to secure the system from corruption by adversaries who are aware of the plaintext message, we define a stronger notion of secure encryption. In particular, we assume that an attacker who is aware of the message $\bfM$, and partially aware of the codeword, $\bfC_{-j} = (C_1,\dots,C_{j-1},C_{j+1},\dots,C_m)$, is statistically incapable of guessing $C_j$ in the absence of knowledge of the key $K$. That is, for any $\bfM, \bfC_{-j}$,
\begin{equation} \label{eqn:enc_given_msg}
\prob{\Phi(\bfM;K) = \bfC \vert \bfM,\bfC_{-j}} \leq \frac{1}{2}, \text{ for any } C_j.
\end{equation}
Note that this criterion indicates that the adversary is unaware of at least $1$ bit of information in the unknown code fragment, despite being aware of the message, i.e.,
\[
H(\bfC \vert \bfM,\bfC_{-j}) \geq 1 \text{ bit}.
\]

\subsection{Distributed Storage Codes}

This work aims to reduce the storage cost for blockchains by using distributed storage codes. Distributed storage codes have been widely studied \cite{DimakisRWS2011, DimakisR2008} in different contexts. In particular, aspects of repair and security, including explicit code constructions have been explored widely \cite{WuDR2007, RashmiSKR2009, ShahRKR2012, PawarRR2011, RawatKSV2014}. In addition, information dispersal algorithms \cite{Rabin1990, CachinT2005} have also considered the question of distributed storage of data. This is a non-exhaustive listing of the existing body of work on distributed storage and most algorithms naturally adapt to the coding scheme defined here. However, we consider the simple form of distributed storage that just divides the data evenly among nodes.

\section{Coding Scheme} \label{sec:coding_scheme}

For this section, assume that at any point of time $t$, there exists a partition $\calP_t$ of the set of peers $[n]$ into sets of size $m$ each. In this work we presume that $n$ is divisible by $m$. Let each set of the partition be referred to as a \emph{zone}. Without loss of generality, the zones are referred to by indices $1,\dots,\tfrac{n}{m}$. At each time $t$, for each peer $i \in [n]$, let $p_t^{(i)} \in [\tfrac{n}{m}]$ be the index that represents the zone that includes peer $i$. We describe the zone allocation scheme in detail in Sec.~\ref{sec:zone_alloc}.

\subsection{Coding Data Block}

\begin{figure}[t]
	\centering
	\includegraphics[scale = 0.5]{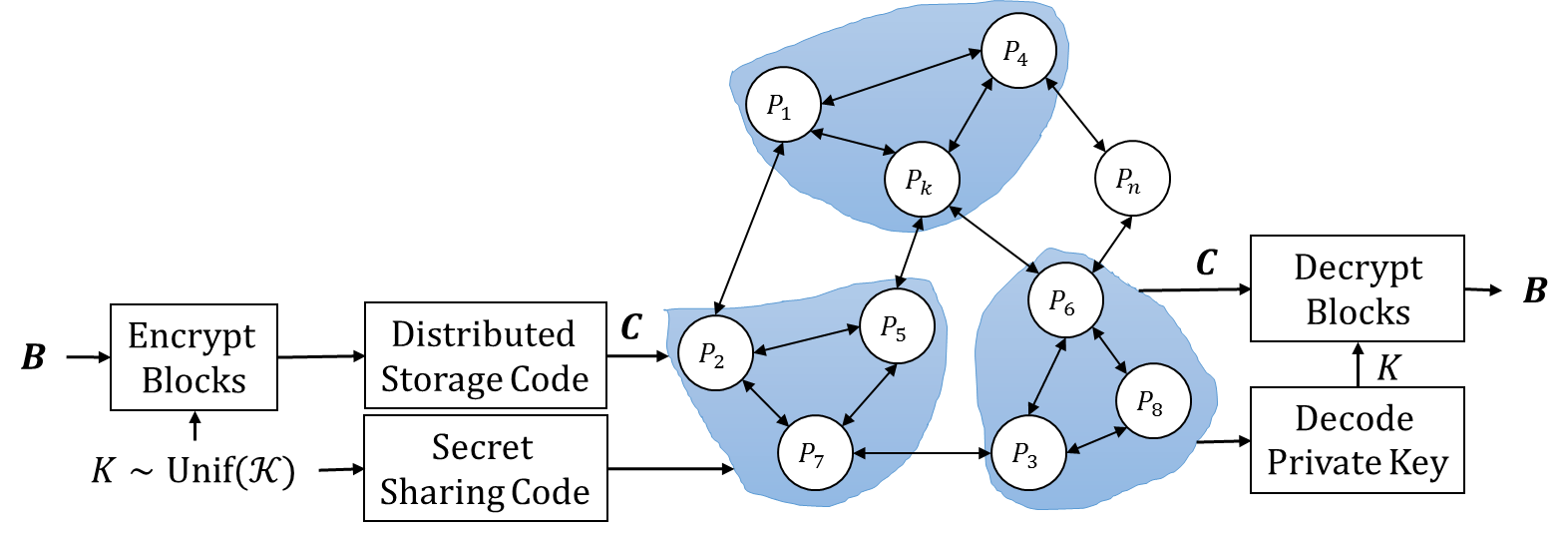}
 	\caption{Encryption and decryption process for a given zone allocation. The shaded regions represent individual zones in the peer network. The data is distributed among peers in each zone and the data from all peers in a zone are required to recover the transaction data.}
  	\label{fig:zone_enc_dec}
\end{figure}

In our coding scheme, a single copy of each data block is stored in a distributed fashion across each zone. Consider the data block $\bfB_t$ corresponding to time $t$. We use a technique inspired by \cite{Krawczyk1994}. First a private key $K$ is generated at each zone and the data block is encrypted using the key. The private key is then stored by the peers in the zone using Shamir's secret key sharing scheme. Finally, the encrypted data block is distributed amongst peers in the zone using a distributed storage scheme. The process involved in storage and recovery of a block, given a zone division is shown in Fig.~\ref{fig:zone_enc_dec}.

The coding scheme is given by Alg.~\ref{alg:block_coding_scheme}.
\begin{algorithm}[t]
  \caption{Coding scheme for data block}
  \label{alg:block_coding_scheme}
  \begin{algorithmic}[t]
   \FOR{$z=1$ to $\tfrac{n}{m}$}
   	\STATE{Generate private key $K^{(z)}_t \sim \text{Unif}(\calK)$}
   	\STATE{Encrypt block with key $K^{(z)}_t$ as $\bfC_t^{(z)} = \Phi(\bfB_t;K^{(z)}_t)$}
   	\STATE{Use a distributed storage code to store $\bfC_t$ among peers in $\{i:p_t^{(i)}=z\}$}
   	\STATE{Use Shamir's $(m,m)$ secret sharing scheme on $K^{(z)}_t$ and distribute the shares $(K_1^{(z)},\dots,K_m^{(z)})$ among peers in the zone}
   \ENDFOR
  \end{algorithmic}
 \end{algorithm}
In this discussion we will assume that the distributed storage scheme just distributes the components of the code vector $\bfC_t$ among the peers in the zone. The theory extends naturally to other distributed storage schemes.

In order to preserve the integrity of the data, we use secure storage for the hash values as well. In particular, at time $t$, each zone $Z \in \calP_t$ stores a secret share of the hash value $H_{t-1}$ generated using Shamir's $(m,m)$ secret sharing scheme.

The storage per transaction per peer is thus given by 
\begin{equation} \label{eqn:dist_storage_cost}
R_s = \frac{1}{m}\log_2 |\calC| + 2\log_2 |\calK| + 2\log_2 p \text{ bits},
\end{equation}
where $|\calC| \geq q$ depending on the encryption scheme. In particular, when the code space of encryption matches the message space, i.e., $|\calC| = q$, the gain in storage cost per transaction per peer is given by
\begin{equation} \label{eqn:storage_gain}
\text{Gain in storage cost} = \tilde{R}_s - R_s = \frac{m-1}{m} \log_2 q - 2\log_2 |\calK| - \log_2 p \text{ bits}.
\end{equation}
Thus, in the typical setting where the size of the private key space is much smaller than the size of the blocks, we have a reduction in the storage cost.

\subsection{Recovery Scheme}

We now describe the algorithm to retrieve a data block $B_t$ in a blockchain system comprising a total of $T$ transactions. The algorithm to recover block $B_t$ is described in Alg.~\ref{alg:recovery_scheme}.

\begin{algorithm}[t]
  \caption{Recovery scheme for data block}
  \label{alg:recovery_scheme}
  \begin{algorithmic}[t]
   \STATE{$\calN \leftarrow [n]$}
   \STATE{Using polynomial interpolation on the shares, compute $K^{(z)}_t$, for all $z \in [\tfrac{n}{m}]$}
   \STATE{Decode blocks $B^{(z)}_t \leftarrow \Psi\pth{\bfC_t^{(z)};K^{(z)}_t}$, for all $z \in [\tfrac{n}{m}]$}
   \IF{$|\{B^{(z)}_t:z \in [\tfrac{n}{m}]\}|>1$}
   	\FOR{$\tau = t$ to $T$}
   		\STATE{Using polynomial interpolation from the hash shares, compute $H^{(z)}_{\tau}$, for all $z \in [\tfrac{n}{m}]$}
   		\STATE{Determine $W^{(z)}_{\tau} = \pth{B^{z}_{\tau}, H^{z}_{\tau-1}}$, for all $z \in [\tfrac{n}{m}]$}
   		\STATE{Determine hash inconsistencies $\calI \leftarrow \sth{i \in [n]: h\pth{W^{(z)}_{\tau}} \neq H^{(z')}_{\tau}, z = p^{(i)}_{\tau}, z' = p^{(i)}_{\tau+1}}$}
   		\STATE{$\calN \leftarrow \calN \backslash \calI$}
   		\IF{$\left|\sth{B^{(p^{(i)}_t)}_t:i \in \calN}\right| = 1$}
   			\STATE{\bf break}
   		\ENDIF
   	\ENDFOR
   \ENDIF
   \RETURN{Majority in $\sth{\sth{B^{(p^{(i)}_t)}_t:i \in \calN }}$}
  \end{algorithmic}
 \end{algorithm}

The recovery algorithm exploits information-theoretic security in the form of the coding scheme, and also invokes the hash-based computational integrity check established in the chain. First, the data blocks are recovered from the distributed, encrypted storage from each zone. In case of a data mismatch, the system inspects the chain for consistency in the hash chain. The system scans the chain for hash values and eliminates peers that have inconsistent hash values. A hash value is said to be inconsistent if the hash value corresponding to the data stored by a node in the previous instance does not match the current hash value.

Through the inconsistency check, the system eliminates some, if not all corrupted peers. Finally, the majority consistent data is returned. In practice, the consistency check along the hash chain can be limited to a finite number of blocks to reduce computational complexity of recovery.

In the implementation, we presume that all computation necessary for the recovery algorithm is done privately by a black box. In particular, we presume that the peers and clients are not made aware of the code stored at other peers or values stored in other blocks. Specifics of practical implementation of such a black box scheme is beyond the scope of this paper.

\subsection{Feasible Encryption Scheme}

The security of the coding scheme from corruption by active adversaries depends on the encryption scheme used. We first describe the necessary condition on the size of the key space.

\begin{lemma}
A valid encryption scheme satisfying \eqref{eqn:enc_given_code} and \eqref{eqn:enc_given_msg}, has 
\[
|\calK| \geq 2^m.
\]
\end{lemma}
\begin{IEEEproof}
First, by chain rule of entropy,
\begin{align}
H(K, \bfC \vert \bfM) &= H(K) + H(\bfC \vert \bfM,K) = H(K), \label{eqn:det_encryption}
\end{align}
where \eqref{eqn:det_encryption} follows from the fact that the codeword is known given the private key and the message.

Again using the chain rule and \eqref{eqn:det_encryption}, we have
\begin{align}
H(K) &= H(\bfC \vert \bfM) + H(K \vert \bfC, \bfM) \notag \\
&\geq \sum_{j=1}^m H(C_j \vert \bfC_{-j},\bfM) \label{eqn:condn_entropy} \\
&\geq m, \label{eqn:min_ent_K}
\end{align}
where \eqref{eqn:condn_entropy} follows from non-negativity of entropy and the fact that conditioning only reduces entropy. Finally, \eqref{eqn:min_ent_K} follows from the condition \eqref{eqn:enc_given_msg}. Since keys are chosen uniformly at random, the result follows.
\end{IEEEproof}

We now describe an encryption scheme that is order optimal in the size of the private key space upto log factors. Let $\calT$ be the set of all rooted, connected trees defined on $m$ nodes. Then, by Cayley's formula \cite{Durrett2007}, 
\[
|\calT| = m^{(m-1)}.
\]
Let us define the key space by the entropy-coded form of uniform draws of a tree from $\calT$. Hence in the description of the encryption scheme, we presume that given the private key $K$, we are aware of all edges in the tree. Let $V = [m]$ be the nodes of the tree and $v_0$ be the root. Let the parent of a node $i$ in the tree be $\mu_i$.

\begin{algorithm}[t]
  \caption{Encryption scheme}
  \label{alg:encryption_scheme}
  \begin{algorithmic}[t]
	\STATE{$T \leftarrow \text{Unif}(\calT)$, $K\leftarrow \text{Key}(T)$; $\bfb \leftarrow \text{Binom}(n,1/2)$}
	\STATE{Assign peers to vertices, i.e., peer $i$ is assigned to node $\theta_i$}
	\STATE{For all $i \neq v_0$, $\tilde{C}_i \leftarrow B_i \oplus B_{\mu_i}$; flip bits if $b_i = 1$.}	
	\STATE{$\tilde{C}_{v_0} \leftarrow \pth{ \oplus_{j \neq v_0} \tilde{C}_j} \oplus B_{v_0}$}
	\IF{$b_{v_0}=1$}
		\STATE{Flip the bits of $\tilde{C}_{v_0}$}
	\ENDIF
	\STATE{Store $C_i \leftarrow \tilde{C}_{\theta_i}$ at each node $i$ in the zone}
	\STATE{Store $(K,\theta)$ using Shamir's secret sharing at the peers}
	\STATE{Store the peer assignment $\theta_i$ locally at each peer $i$}
  \end{algorithmic}
 \end{algorithm}

\begin{figure}[t]
	\centering
	\includegraphics[scale = 0.4]{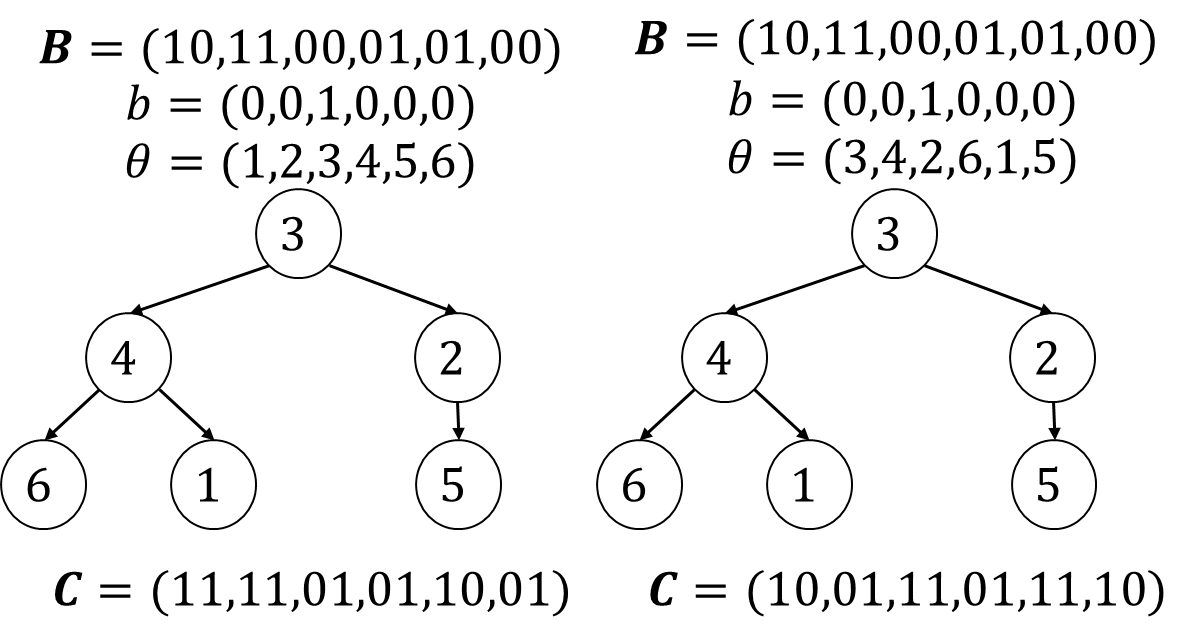}
 	\caption{Encryption examples for a zone with six peers. The data block, parameters, tree structure, and corresponding codes are shown. The two cases consider the same rooted tree with varying peer assignments. The corresponding change in the code is shown.}
  	\label{fig:enc_tree}
\end{figure}

Consider the encryption function given in Alg.~\ref{alg:encryption_scheme}. The encryption algorithm proceeds by first selecting a rooted, connected tree uniformly at random on $m$ nodes. Then, each peer is assigned to a particular node of the tree. For each node other than the root, the codeword is created as the modulo $2$ sum of the corresponding data block and that corresponding to the parent. Finally, the root is encrypted as the modulo $2$ sum of all codewords at other nodes and the corresponding data block. The bits stored at the root node are flipped with probability half. The encryption scheme for a sample data block is shown in Fig.~\ref{fig:enc_tree}. We refer to Alg.~\ref{alg:encryption_scheme} as $\Phi$ from here on.

The decryption of the stored code is as given in Alg.~\ref{alg:decryption_scheme}. That is, we first determine the private key, i.e., the rooted tree structure, the bit, and peer assignments. Then we decrypt the root node by using the codewords at other peers. Then we sequentially recover the other blocks by using the plain text message at the parent node.

\begin{algorithm}[t]
  \caption{Decryption scheme}
  \label{alg:decryption_scheme}
  \begin{algorithmic}[t]
	\STATE{Use polynomial interpolation to recover $(K,\bfb,\theta)$}
	\STATE{Define $\tilde{\theta}_i \leftarrow j$ if $\theta_j = i$}
	\STATE{Flip the bits of $C_{\tilde{\theta}_{v_0}}$, if $b_{v_0}=1$}
	\STATE{$B_{v_0} \leftarrow C_{\tilde{\theta}_{v_0}} \oplus_{j \neq \tilde{\theta}_{v_0}} C_j$}
	\STATE{For all $i \in [n]\backslash \{v_0\}$, flip bits of $\bfC_i$ if $b_i = 1$}
	\STATE{Iteratively compute $B_i \leftarrow C_{\tilde{\theta}_i} \oplus B_{\mu_i}$  for all $i \neq v_0$}
	\RETURN{$\bfB$}
  \end{algorithmic}
 \end{algorithm}
 
\begin{lemma}
The encryption scheme $\Phi$ satisfies \eqref{eqn:enc_given_code} and \eqref{eqn:enc_given_msg}.
\end{lemma}
\begin{IEEEproof}
Validity of \eqref{eqn:enc_given_code} follows directly from the definition of the encryption scheme as the message is not recoverable from just the encryption. 

To check the validity of \eqref{eqn:enc_given_msg}, note that given the data at all peers other than one node, the adversary is unaware of the parent of the missing node. Since this is uniformly likely, the probability that the adversary can guess the encrypted data is at most $1/2$, with the maximum being if the root is not recovered.
\end{IEEEproof}

\begin{lemma} \label{lem:storage_cost_lemma}
The storage cost per peer per transaction under $(\Phi,\Psi)$ is
\begin{equation}
R_s(\Phi,\Psi) = \frac{1}{m} \log_2 q + 2 m \log_2 m + 2 \log_2 p + 1 \text{ bits}
\end{equation}
\end{lemma}
\begin{IEEEproof}
First note that $\calC = \bbF_q$. Next, the number of rooted, connected trees on $m$ nodes is given by Cayley's formula as $m^{(m-1)}$. The peer assignments can be stored locally and so cost only $\log_2 m$ bits per node per transaction. Thus, the result follows.
\end{IEEEproof}
From Lemma \ref{lem:storage_cost_lemma}, we can see that the encryption scheme guarantees order-optimal storage cost per peer per transaction up to $\log$ factor in the size of the key space. The security of the encrypted data can be enhanced by increasing the inter-data dependency by using directed acyclic graphs (DAGs) with bounded in-degree in place of the rooted tree. Then, the size of storage for the private key increases by a constant multiple.

\subsection{Individual Block Corruption}

We now establish the security guarantees of individual blocks in each zone from active adversaries. First, consider an adversary who is aware of the hash value $H_t$ and wishes to alter it to $H_t'$.
\begin{lemma} \label{lem:hash_corr}
Say an adversary, aware of the hash value $H_t$ and the peers in a zone $z$, wishes to alter the value stored in the zone to $H_t'$. Then, the probability of successful corruption of such a system when at least one peer is honest, is $O(1/q)$.
\end{lemma}
\begin{IEEEproof}
Assume the adversary knows the secret shares of $k-1$ peers in the zone. Since the adversary is also aware of $H_t = a_0$, the adversary is aware of the coding scheme through polynomial interpolation. However, since the final peer is honest, the adversary is unaware of the secret share stored here. Hence the result follows.
\end{IEEEproof}
This indicates that in order to corrupt a hash value, the adversary practically needs to corrupt all nodes in the zone.

To understand corruption of data blocks, we first consider the probability of successful corruption of a zone without corrupting all peers of the zone.
\begin{lemma} \label{lem:zonal_corruption}
Consider an adversary, aware of the plain text $\bfB$ and the peers in a zone. If the adversary corrupts $c < m$ peers of the zone, then the probability that the adversary can alter the data to $\bfB'$ is at most $\frac{c^2}{m^2}$, i.e.,
\begin{equation}
\prob{\bfB \rightarrow \bfB' \text{ in zone } z} \leq \frac{c}{m}, \text{ for all } \bfB \neq \bfB', z \in [\tfrac{n}{m}].
\end{equation}
\end{lemma}
\begin{IEEEproof}
For ease, let us assume that $\theta_i = i$ for all $i \in [m]$. From the construction of the encryption scheme, we note that if $B_i \rightarrow B_i'$ is to be performed, then all blocks in the subtree rooted at $i$ are to be altered as well. Further, for any change in the block contents, the root is also to be altered.

Thus, a successful corruption is possible only if all nodes in the subtree and the root have been corrupted. However, the adversary can only corrupt the peers at random and has no information regarding the structure of the tree. Thus, the simplest corruption is one that alters only a leaf and the root. Thus, the probability of corruption can be bounded by the probability of corrupting those two particular nodes as follows:
\begin{align}
\prob{\bfB \rightarrow \bfB' \text{ in zone } z} &\leq \prob{\text{Selecting a leaf and root in } c \text{ draws without replacement from } [m]} \notag \\
&= \binom{m-2}{c-2}/\binom{m}{c} \notag \\
&= \frac{c(c-1)}{m(m-1)} \leq \frac{c^2}{m^2}.
\end{align}
\end{IEEEproof}
A consistent corruption of a transaction by an active adversary however requires corruption of at least $\tfrac{n}{2m}$ zones. 

\begin{theorem} \label{thm:consistent_corruption}
Consider an active adversary who corrupts $c_1,\dots,c_{n/2m}$ peers respectively in $n/2m$ zones. Then, the probability of successful corruption across the set of all peers is 
\begin{equation}
\prob{\text{Successful corruption } \bfB\rightarrow\bfB'} \leq \exp\pth{\frac{n}{m}\log\pth{\frac{2\sum_{i=1}^{n/2m}c_i}{n}}}, \text{ for all } \bfB \neq \bfB'. 
\end{equation}
\end{theorem}
\begin{IEEEproof}
From Lemma \ref{lem:zonal_corruption} and independence of the encryption across zones, we have
\begin{align}
\prob{\text{Successful consistent corruption } \bfB\rightarrow\bfB'} &\leq \prod_{i=1}^{n/2m} \frac{c_i^2}{m^2} \notag \\
&= \exp\pth{2 \sum_{i=1}^{n/2m} \log c_i - 2 \frac{n}{2m} \log m} \notag \\
&\leq \exp\pth{ \frac{n}{m} \log \pth{\tfrac{2\sum_{i=1}^{n/2m} c_i}{n}}}, \label{eqn:AM-GM} 
\end{align}
where \eqref{eqn:AM-GM} follows from the arithmetic-geometric mean inequality.
\end{IEEEproof}
Note that 
\[
\sum_{i=1}^{n/2m} c_i \leq \frac{n}{2},
\]
and thus the upper bound on successful corruption decays with the size of the peer network if less than half the network is corrupted. From Thm.~\ref{thm:consistent_corruption}, we immediately get the following corollary.
\begin{corollary} \label{cor:total_corr}
If an adversary wishes to corrupt a data block with probability at least $1-\epsilon$, for some $\epsilon > 0$, then the total number of nodes to be corrupted satisfies
\begin{equation}
\sum_{i=1}^{n/2m} c_i \geq \frac{n}{2} (1-\epsilon)^{\tfrac{m}{n}}.
\end{equation}
\end{corollary}
Cor.~\ref{cor:total_corr} indicates that when the network size is large, the adversary practically needs to corrupt at least half the network to have the necessary probability of successful corruption. Thus we observe that for a fixed zone division, the distributed storage system loses an arbitrarily small amount of data integrity as compared to the conventional scheme. 

In Sec.~\ref{sec:zone_alloc}, we introduce a dynamic zone allocation scheme to divide the peer network into zones for different time slots. We show that varying the zone allocation patterns over time appropriately yields even better data integrity. 

With regard to denial of service attacks, since it is infeasible to recover the data block without the share of any peer in the zone, the system is capable of handling upto $1$ denial of service attack per zone. That is, the system can tolerate a total of $n/m$ denial of service attacks.

\subsection{Data Confidentiality}

We earlier stated that the two aspects of security needed in blockchain systems are data integrity and confidentiality. We addressed the question of data integrity in the previous subsection. We now consider confidentiality of transaction data.

Consider the situation where a peer $i$ in a zone is compromised. That is, an external adversary receives the data stored by the peer for one particular slot. This includes the secret share of the private key $K_i$, the encrypted block data $C_i$, and secret share corresponding to the hash of the previous block.

From Shamir's secret sharing scheme, we know that knowledge of $K_i$ gives no information regarding the actual private key $K$. Thus, the adversary has no information on the rooted tree used for encryption.

We know that the transaction data are chosen uniformly at random. Since the adversary is unaware of the relation of the nodes to one another, from the encryption scheme defined, we know that given the entire encrypted data $\bfC$, the probability of recovering the block $\bfB$ is uniformly distributed on the set of all possible combinations obtained for all possible tree configurations. That is, each possibly rooted tree yields a potential candidate for the transaction data.

This observation implies that
\begin{equation}
H(\bfB \vert \bfC) \leq H(K, \bfB \vert \bfC) = H(K) + H(\bfB \vert K, \bfC) = m \log_2 m.
\end{equation}
We know that the entropy of the transaction block is actually $H(\bfB) = \log_2 q > m \log_2 m$. That is, the adversary does learn the transaction data partially and has a smaller set of candidates in comparison to the set of all possible values, given the entire codeword.

However, in the presence of just $C_i$, the adversary has no way to determine any of the other stored data, nor does it have any information on the position of this part of the code in the underlying transaction block. Thus, local leaks reveal very little information regarding the transaction block.

Thus, we observe that the coding scheme also ensures a high degree of confidentiality in case of data leaks from up to $m-1$ peers in a zone.

\section{Dynamic Zone Allocation} \label{sec:zone_alloc}

In the definition of the coding and recovery schemes, we presumed the existence of a zone allocation strategy over time. Here we make it explicit. 

Cor.~\ref{cor:total_corr} and Lemma~\ref{lem:hash_corr} highlighted the fact that the distributed secure encoding process ensures that corrupting a transaction block or a hash requires an adversary to corrupt all peers in the zone. This fact can be exploited to ensure that with each transaction following the transaction to be corrupted, the client would need to corrupt an increasing set of peers to maintain a consistent version of the corrupted chain.

\begin{figure}[t]
	\centering
	\includegraphics[scale = 0.5]{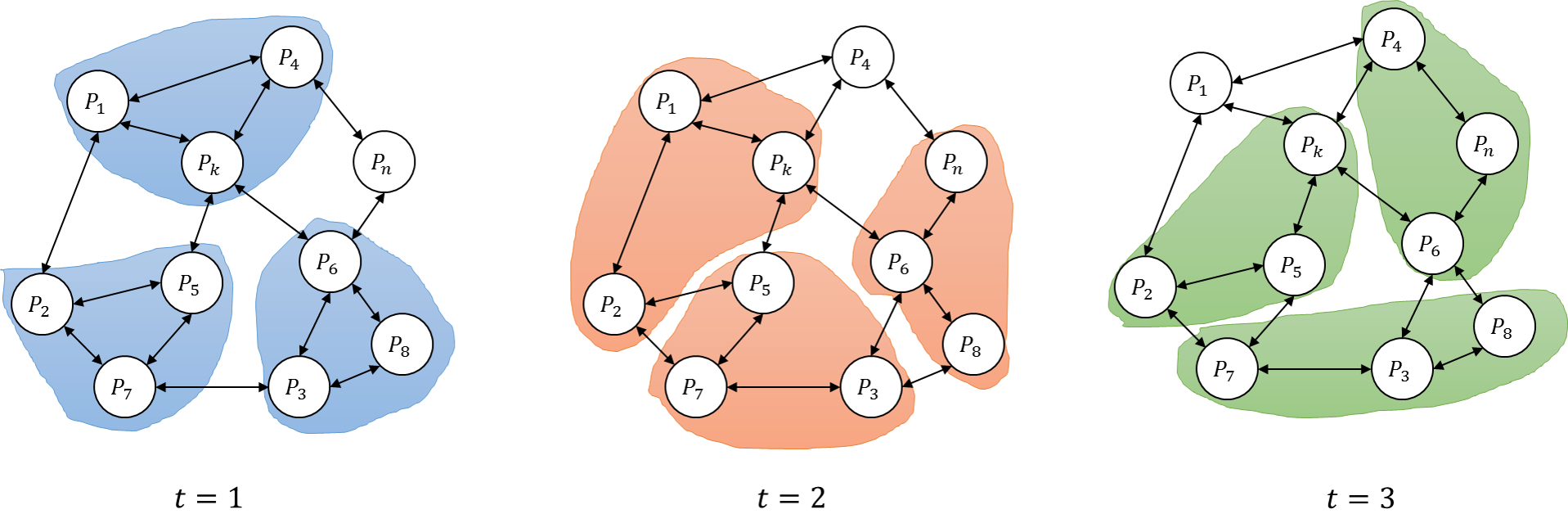}
 	\caption{Dynamic Zone Allocation over Time: Iterate the zone allocation patterns among the peers so that increasing number of peers need to be corrupted to maintain a consistent chain structure.}
  	\label{fig:network_distrib_dynamic}
\end{figure}

In particular, let us assume a blockchain in the following state
\[
(H_0,\bfB_1)-(H_1,\bfB_2)-\dots-(H_{t-1},\bfB_t).
\]
Let us assume without loss of generality that an adversary wishes to corrupt the transaction entry $\bfB_1$ to $\bfB_1'$. The validated, consistent version of such a corrupted chain would look like
\[
(H_0,\bfB_1')-(H_1',\bfB_2)-\dots-(H_{t-1}',\bfB_t).
\]
If the zone segmentation used for the encoding process is static, then the adversary can easily maintain such a corrupted chain at half the peers to validate its claim. If each peer is paired with varying sets of peers across blocks, then, for sufficiently large $t$, each corrupted peer eventually pairs with an uncorrupted peer. 

Let us assume that this occurs for a set of corrupted peers at slot $\tau$. Then, in order to successfully corrupt the hash $H_{\tau - 1}$ to $H_{\tau - 1}'$, the adversary would need to corrupt the rest of the uncorrupted peers in the new zone. On the other hand, if the client does not corrupt these nodes, then the hash value remains unaltered indicating the inconsistencies of the corrupted peers. 

Thus, it is evident that if the zones are sufficiently well distributed, corrupting a single transaction would eventually require corruption of the entire network, and not just a majority. A sample allocation scheme is shown in Fig.~\ref{fig:network_distrib_dynamic}. 

However, the total number of feasible zone allocations is given by
\begin{align}
\text{No. of zone allocations} &= \frac{n!}{(m!)^{\tfrac{n}{m}}} \approx \frac{\sqrt{2\pi n}}{\pth{\sqrt{2\pi m}}^{\tfrac{n}{m}}} \pth{\frac{n}{m}}^n,
\end{align}
which increases exponentially with the number of peers and is monotonically decreasing in the zone size $m$. This indicates that naive deterministic cycling through this set of all possible zone allocations is practically infeasible.

To ensure that every uncorrupted peer is eventually grouped with a corrupted peer, we essentially need to ensure that every peer is eventually grouped with every other peer. Further, the blockchain system needs to ensure uniform security for every transaction and to this end, the allocation process should also be fair.

In order to better understand the zone allocation strategy, we first study a combinatorial problem.

\subsection{$K$-way Handshake Problem}

Consider a group of $n$ people. At each slot of time, the people are to be grouped into sets of size $m$. A peer gets acquainted with all other peers in the group whom they have not met before. The problem can thus be viewed as an $m$-way handshake between people.
\begin{lemma}
The minimum number of slots required for every peer to shake hands with every other peer is $\tfrac{n-1}{m-1}$.
\end{lemma}
\begin{IEEEproof}
At any slot, a peer meets at most $m-1$ new peers. Thus the lower bound follows.
\end{IEEEproof}

\begin{remark}
Note that each such grouping of nodes constitutes a matching ($1$-factor) of an $m$-uniform complete hypergraph on $n$ nodes, $K_m^n$. Baranyai's theorem \cite{Baranyai1979} states that if $n$ is divisible by $m$, then, there exists a decomposition of $K_m^n$ into $\binom{n-1}{m-1}$ $1$-factors. However, we do not require every hyperedge to be covered by the allocation scheme, but only for every node to be grouped with every other node eventually. 

Note that for $m=2$, it shows that we can decompose a graph into $n-1$ different matchings. In this case, the handshake problem is the same as the decomposition of the graph into matchings. Thus Baranyai's theorem in this case gives us the exact number of slots to solve the problem.
\end{remark}

We use the tightness observed for the $2$-way handshake problem to design a strategy to assign the peers in zones. Let $n' = \tfrac{n}{m}$. Partition the peers into $2n'$ sets, each containing $m/2$ peers. Let these sets be given by $\nu_1,\dots,\nu_{2n'}$. Then, we can use matchings of $K_{2n'}$ to perform the zone allocation.

\begin{algorithm}[t]
  \caption{Dynamic Zone Allocation Strategy}
  \label{alg:zone_alloc}
  \begin{algorithmic}[t]
	\STATE{Let $\nu_2\dots,\nu_{2n'}$ be the vertices of a $2n'-1$ regular polygon, and $\nu_1$ its center}
	\FOR {$i=2$ to $2n'$}	
		\STATE{Let $L$ be the line passing through $\nu_1$ and $\nu_i$}
		\STATE{$M \leftarrow \sth{(\nu_j,\nu_k):\text{ line through } \nu_j,\nu_k \text{ is perpendicular to } L}$}
		\STATE{$M \leftarrow M \cup \{(\nu_1,\nu_i)\}$}
		\STATE{Construct zones as $\{\nu_j\cup\nu_k : (\nu_j,\nu_k) \in M\}$}
	\ENDFOR
	\STATE{{\bf restart} for loop}
  \end{algorithmic}
 \end{algorithm}

\begin{figure}[t]
	\centering
	\includegraphics[scale = 0.3]{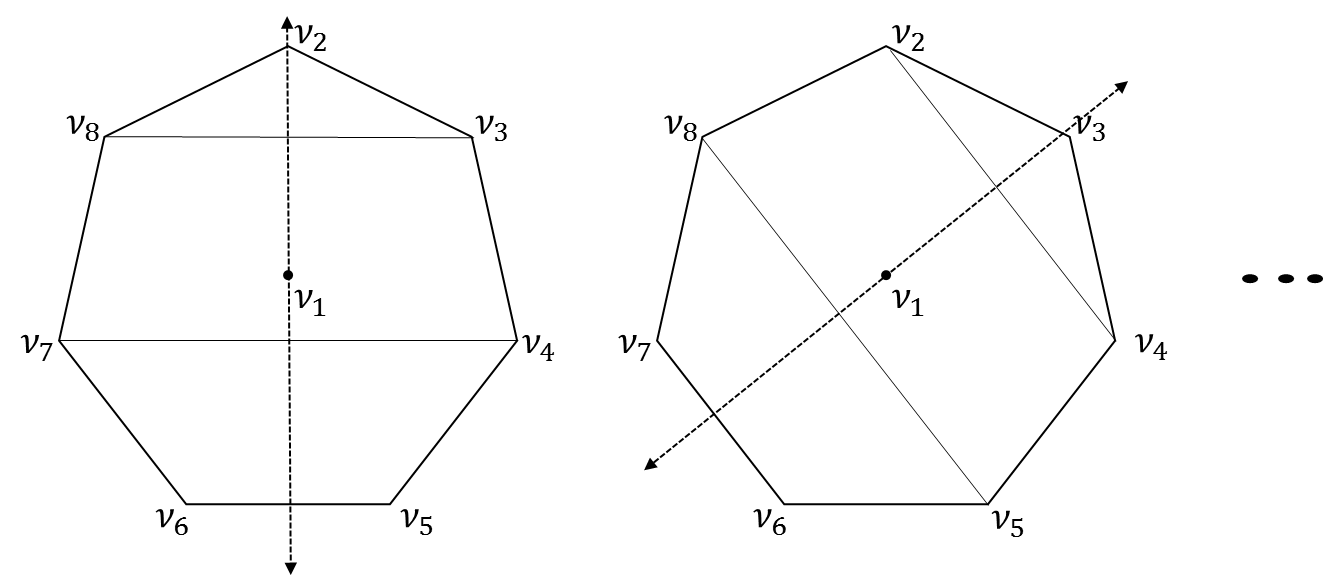}
 	\caption{Dynamic Zone Allocation strategy when $n = 4m$. The zone allocation scheme cycles through matchings of the complete graph by viewing them in the form of the regular polygon.}
  	\label{fig:zone_alloc}
\end{figure}

Consider Alg.~\ref{alg:zone_alloc}. The algorithm provides a constructive method to create zones such that all peers are grouped with each other over time. The functioning of the algorithm is as in Fig.~\ref{fig:zone_alloc}.

\begin{lemma} \label{lem:achievable_coverage}
The number of slots required for every peer to be grouped with every other peer is $2n' - 1$. 
\end{lemma}
\begin{IEEEproof}
The result follows directly from the cyclic decomposition and Baranyai's theorem for $m=2$.
\end{IEEEproof}
We see that the scheme matches the lower bound on the number of slots for coverage in the order sense. Thus we consider this allocation strategy in the following discussion. In addition to the order optimality, the method is also fair in its implementation to all transactions over time.

\subsection{Security Enhancement}

From Alg.~\ref{alg:recovery_scheme}, we know that inconsistent peers are removed from consideration for data recovery. While Lemma~\ref{lem:achievable_coverage} guarantees coverage in $2\tfrac{n}{m}$ slots, we are in fact interested in the number of slots for all uncorrupted peers to be paired with corrupt peers. We now give insight into the rate at which this happens.

We know that an adversary who wishes to corrupt a block corrupts at least $n/2$ nodes originally. 
\begin{lemma}
Under the cyclic zone allocation strategy, the adversary needs to corrupt at least $m$ new nodes with each new transaction to establish a consistent corruption.
\end{lemma}
\begin{IEEEproof}
From the cyclic zone allocation strategy and the pigeonhole principle, we note that with each slot, at least two honest nodes in the graph are paired with corrupt nodes. Thus the result follows.
\end{IEEEproof}
Naturally this implies that in the worst case, with $\tfrac{n}{2m}$ transactions, the data becomes completely secure in the network. That is, only a corruption of \emph{all peers} (not just a majority) leads to a consistent corruption of the transaction.

\section{Data Recovery and Repair} \label{sec:repair_recovery}

\subsection{Recovery Cost}

As highlighted in the scheme, the recovery process of a transaction block or hash value requires the participation of all peers in the zones. However, practical systems often have peers that are temporarily inactive or undergo data failure. In such contexts the recovery of the data from the corresponding zone becomes infeasible.

Thus it is of interest to know the probability that it may not be feasible to recover an old transaction at any time slot. Consider a simple model wherein the probability that a peer is inactive in a slot is $\rho$, and peer activity across slots and peers is independent and identically distributed.

\begin{theorem} \label{thm:recovery}
For any $\delta > 0$, probability of successful recovery of a data block at any time slot is at least $1-\delta$ if and only if $m = \Theta(\log n)$ 
\end{theorem}
\begin{IEEEproof}
First, the probability that the data stored can be recovered at any time slot is bounded according to the union bound as follows:
\begin{align}
\prob{\text{Recovery}} &= \prob{\text{there exists a zone with all active peers}} \notag \\
&\leq \frac{n}{m}(1-\rho)^m. \notag
\end{align}
Thus, to guarantee a recovery probability of at least $1-\delta$, we need
\begin{align}
\frac{n}{m}(1-\rho)^m &\geq 1 - \delta \label{eqn:rec_prob_necc1} \\
\implies m &\leq \frac{1}{\log \pth{\tfrac{1}{1-\rho}}} \pth{\log n - \log (1-\delta)}. \label{eqn:rec_prob_necc2}
\end{align}

Next, to obtain sufficient conditions on the size of the zones, note that
\begin{align}
\prob{\text{Failure}} &= \prob{\text{at least one peer in each zone is inactive}} \notag \\
&= \pth{1 - (1-\rho)^m}^{\tfrac{n}{m}} \leq \exp\pth{-(1-\rho)^m \frac{n}{m}}, \notag
\end{align}
where the last inequality follows from the fact that $1-x \leq \exp(-x)$. Hence a sufficient condition for guaranteeing an error probability of less than $\delta$ is
\begin{align}
\exp\pth{-(1-\rho)^m \frac{n}{m}} &\leq \delta \notag \\
\implies m &\leq \frac{1}{\log \pth{\tfrac{1}{1-\rho}}} \pth{\log n - \log \log (1/\delta)}. \label{eqn:rec_prob_suff}
\end{align}
Thus the result follows.
\end{IEEEproof}

Thm.~\ref{thm:recovery} indicates that the zone sizes have to be of the order of $\log n$ to guarantee a required probability of recovery in one slot when node failures are possible.

\subsection{Data Repair}

The distributed secure storage ensures that individual entries stored at each peer can not be recovered from the knowledge of other entries in the zone. Thus it is not feasible to repair nodes locally within a zone. However, it suffices to substitute a set of bits such that the code structure is retained.

A node failure indicates that the private key is lost. Thus repairing a node involves recoding the entire zone using data from a neighboring zone. Thus owing to the encryption, it is difficult to repair nodes upon failure. 

Thus, the transaction data is completely lost if one peer from every zone undergoes failure. That is, the system can handle up to $n/m$ node failures. This again emphasizes the need to ensure that $m$ is small in comparison to $n$.

\section{Computation and Energy} \label{sec:energy_redn}

As stated in Sec.~\ref{sec:sys_model_b}, conventional blockchain systems incorporate constraints on the hash values of transactions over time to enhance the security of data blocks. Enforcing such constraints however translates to higher computational load on the block validation (mining) process.

Let us elaborate with the example of the bitcoin. The bitcoin network employs the SHA-256 algorithm for generating the hash values \cite{GilbertH2004}. At each time slot, the network chooses a difficulty level at random from the set of hash values, $\bbF_p$. The role of the miner in the network is to append an appropriate nonce to the block such that the corresponding hash value satisfies the difficulty level for the slot. The nonce-appended block and the generated hash value are then added to the blockchain. To be precise, the hash is applied twice in bitcoin. First, the block is hashed and then the nonce is appended to this version and rehashed. However, for simplicity, we just presume that the nonce is added directly to the hash.

At any slot $t$, let the transaction block be $B_t$ and previous hash $H_{t-1}$. Let the set of all possible nonce values be $\calN = \{0,1\}^{b}$, where $b$ is the number of bits appended to the transaction block, and is chosen sufficiently large such that the hash criterion will be satisfied for at least one value. Let the nonce added by the miner be $N_t$. Then, the miner searches for $N_t \in \calN$ such that
\[
h(H_{t-1},N_t,B_t) \in \calH_t,
\]
where $\calH_t \subseteq \bbF_p$ is the set of permissible hash values at time $t$ as determined by the difficulty level.

\begin{algorithm}[t]
  \caption{Bitcoin Mining Algorithm}
  \label{alg:bitcoin_mining}
  \begin{algorithmic}[t]
	\STATE{{\bf input:} previous block data $W_{t}$, acceptable hash value set $\calH_t$}
	\STATE{{\bf output:} nonce-appended data $\tilde{W}_t$, hash value $H_t$}
	\FORALL{$N_t \in \calN$}
		\STATE{$\tilde{W}_t \leftarrow (N_t,W_t)$}
		\STATE{$H_t \leftarrow h(\tilde{W}_t)$}
		\IF{$H_t \in \calH_t$}
			\STATE{\bf break}
		\ENDIF
	\ENDFOR
	\STATE{{\bf return} $\tilde{W}_t, H_t$}
  \end{algorithmic}
 \end{algorithm}

However, due to the pre-image resistance of the cryptographic hash function, the miner is not capable of directly estimating the nonce, given the data block and difficulty level. This in conjunction with collision resistance of the hash function implies that the best way for the miner to find a nonce that corresponds to a hash value in the desired range is to perform brute-force search. Thus, the typical miner in the bitcoin network adopts Alg.~\ref{alg:bitcoin_mining} to add a transaction to the blockchain. The brute-force search for nonce value results in a large computational cost which in turn results in an increase in the consumption of electricity as reflected in the hash rate as was shown in Fig.~\ref{fig:hash_rate}.

The imposition of the constraints on the hash values for each block enhances the integrity of data blocks. Consider an adversary who corrupts a data block. Establishing consistency in the chain following such a corruption requires not only recomputing a hash value, but doing so such that the hash values adhere to the constraints. Each such computation of a hash value is computationally expensive. Thus, the security established through the use of such constraints is essentially equivalent to preventing the addition of incorrect transactions to the chain, as is to be prevented by expensive mining.

We now use the notion of an ideal cryptographic hash function to quantify the expected computational cost. Let us again presume that the message space is bounded and given by $\calM$. If $\bfM \sim \text{Unif}(\calM)$ and $h(\cdot)$ is an ideal cryptographic hash function, then $H = h(\bfM)$ is also distributed uniformly in the space of hash values $\calH$. Thus, extending from this notion, for any $\calH_t \subset \calH$, the set of values $\calM_t \subset \calM$ such that $h(\bfM) \in \calH_t$ for any $\bfM \in \calM_t$ has a volume proportional to the size of the hash subset, i.e., $|\calM_t| \propto |\calH_t|$.

We know that the mining algorithm exhaustively searches for a nonce value that matches the required hash constraints. Now, if the hash value is chosen at random, and the transaction block is given to be uniformly random, then searching for the nonce by brute-force is equivalent, in average to the experiment of drawing balls from an urn without replacement.

\begin{theorem} \label{thm:det_unif_relation}
Let $q = |\calN|$, $|\calH_t| = p'$ be given, and let $\calH_t$ be chosen at uniformly at random from the set of all subsets of $\calH$ with size $p'$. Let $M(\bfW_t,\calH_t)$ be the number of steps taken by Alg.~\ref{alg:bitcoin_mining} to find a valid hash. If $1 \ll p' \ll p$, and $b \geq p'$ is sufficiently large, then
\begin{equation} \label{eqn:det_unif_relation}
\expect{M(\bfW_t,\calH_t) \vert \bfW_t,|\calH_t| = p'} = \expect{M'\pth{\tfrac{p'}{p}q,\pth{1-\tfrac{p'}{p}}q}},
\end{equation}
where $M'(a,b)$ is the number of turns it takes, drawing without replacement from an urn containing $a_1$ blue and $a_2$ red balls, to get the first blue ball.
\end{theorem}
\begin{IEEEproof}
First, we know that the volume of permissible nonce values is proportional to the size of the hash subset, i.e., is $c p'$, for some constant $c$. Any message from this set is uniformly likely as the corresponding constraint set is drawn uniformly. Further, we know that for $p' = p$, any nonce is permissible and so $cp = q$.

Thus, we essentially are drawing without replacement as the brute-force search does not resample candidates, and we are drawing till we get an element from the set of size $\tfrac{p'}{p}q$ in a set of size $q$. This is essentially the same as would be the case for sampling without replacement from the urn. Since the uniformity in trials holds only on average, the result follows.
\end{IEEEproof}
\begin{corollary}
The expected computational cost of the mining algorithm is given by 
\[
\expect{M(\bfW_t,\calH_t) \vert \bfW_t,|\calH_t| = p'} = O\pth{\frac{p}{p'}},
\]
for any $1 \ll p' \ll p$, and $b \geq p'$ sufficiently large.
\end{corollary}
\begin{IEEEproof}
The result follows directly from Thm.~\ref{thm:det_unif_relation} and the expected number of draws for the sampling without replacement experiment.
\end{IEEEproof}
In practical systems, a moderate range of values of $p'$ are used as a very small subset might not be satisfied with a bounded nonce set and a large subset leads to loss of data integrity, as almost any value works. Thus practical systems choose values in a range wherein it is sufficiently difficult to generate viable candidates satisfying the hash constraint. 

Thus we note that the average computational cost can be very high in bitcoin-like systems. In particular, the bitcoin uses $256$ bit hash values. That is, $\log_2 p = 256$, while $\log_2 q = 32$. This in turn implies that the potential candidate space to search is very large thereby leading to increased energy consumption.

On the other hand, in our proposed dynamic distributed storage scheme, the hash values computed are not restricted by these constraints. Instead we enhance data integrity in the blockchain using information-theoretic secrecy instead of a computational one. This way, computing hash values and the corresponding secret shares is computationally inexpensive. Further, we have shown that the integrity is guaranteed through the consistency checks. Thus, the scheme defined here guarantees a reduction in the computational cost without compromising on security.

\section{Conclusion}

In this work, we considered fundamental questions that hinder the scaling of conventional blockchain systems. We addressed the issue of increasing storage costs associated with large-scale blockchain systems. Using a novel combination of secret key sharing, encryption, and distributed storage, we developed coding schemes that were information-theoretically secure and reduced the storage to a fraction of the original load. We also showed that the resulting storage cost was optimal up to log factors in the size of the private key.

We showed that the distributed storage code ensures both data integrity and confidentiality, with very little loss in comparison to the conventional schemes. We then used a dynamic segmentation scheme to enhance data integrity. We described a zone allocation strategies using decomposition of a complete graph into matchings that was order optimal in the time for complete coverage.

We also investigated the energy requirements of conventional bitcoin-like blockchain systems by studying the computational complexity associated with establishing the hashchain. We highlighted the need for simple hashing schemes and showed that our coding scheme guarantees a significant reduction in energy consumption.

Notwithstanding these benefits, the coding scheme hinges centrally on distributing the transaction data securely among a set of peers. Thus, the recovery and repair costs associated with the scheme are higher than conventional systems. In  particular, through probabilistic recovery arguments, we established order optimal necessary and sufficient conditions on the size of the zones in our construction. Further, we also highlighted the fact that local repair is infeasible in this scheme and requires explicit recomputation of the codes corresponding to the zone.

Additionally, the encoding and recovery schemes presumed the feasibility of performing them using black box methods, such that the peers do not learn the codes pertaining to other peers. We did not attempt to define explicit methods of constructing such black box schemes in this work, and this would be the focus of future research. 

Another aspect to be addressed here is the computational protocols and the related costs to be used to distributed the codewords in the peer-to-peer network. Since the data stored by individual nodes varies, owing to the dynamic distributed storage scheme, it is essential to communicate the individual blocks in a point-to-point manner, rather than as a broadcast, as is done in conventional blockchain systems. This process naturally implies an increased communication cost in the network. Since this is closely related to the code construction methodology, this aspect is also to be considered in detail in future research.

This work establishes the feasibility of enhancing blockchain performance and reducing associated costs through novel use of coding-theoretic techniques. We believe such methods enhance the ability of blockchain systems to scale to address practical applications in a variety of industries.

\section*{Acknowledgment}

The authors would like to thank Prof. Andrew Miller for references and useful discussions on blockchain systems.

\bibliographystyle{IEEEtran}
\bibliography{abrv,conf_abrv,lrv_lib}

\end{document}